\documentclass[12pt,letter]{article}

\usepackage{amsfonts,amssymb,amsmath,amsthm,bm}
\usepackage{latexsym,url,color,enumerate}
\usepackage{graphicx}
\usepackage{kbordermatrix}
\usepackage{float}
\usepackage[normalem]{ulem}

\newtheorem{theorem}{Theorem}

\newtheorem{lemma}{Lemma}

\newtheorem{example}{Example}
\newtheorem{fact}{Fact}
\newtheorem{remark}{Remark}

\newcommand{\footremember}[2]{%
    \footnote{#2}
    \newcounter{#1}
    \setcounter{#1}{\value{footnote}}%
}
\newcommand{\footrecall}[1]{%
    \footnotemark[\value{#1}]%
}

\newcommand\nc\newcommand
\nc\bfa{{\boldsymbol a}}\nc\bfA{{\bf A}}\nc\cA{{\mathcal A}}
\nc\bfb{{\boldsymbol b}}\nc\bfB{{\bf B}}\nc\cB{{\mathcal B}}
\nc\bfc{{\boldsymbol c}}\nc\bfC{{\bf C}}\nc\cC{{\mathcal C}}
\nc\bfd{{\boldsymbol d}}\nc\bfD{{\bf D}}\nc\cD{{\mathcal D}}
\nc\bfe{{\boldsymbol e}}\nc\bfE{{\bf E}}\nc\cE{{\mathcal E}}
\nc\bff{{\boldsymbol f}}\nc\bfF{{\bf F}}\nc\cF{{\mathcal F}}
\nc\bfg{{\boldsymbol g}}\nc\bfG{{\bf G}}\nc\cG{{\mathcal G}}
\nc\bfh{{\boldsymbol h}}\nc\bfH{{\bf H}}\nc\cH{{\mathcal H}}
\nc\bfi{{\boldsymbol i}}\nc\bfI{{\bf I}}\nc\cI{{\mathcal I}}
\nc\bfj{{\boldsymbol j}}\nc\bfJ{{\bf J}}\nc\cJ{{\mathcal J}}
\nc\bfk{{\boldsymbol k}}\nc\bfK{{\bf K}}\nc\cK{{\mathcal K}}
\nc\bfl{{\boldsymbol l}}\nc\bfL{{\bf L}}\nc\cL{{\mathcal L}}
\nc\bfm{{\boldsymbol m}}\nc\bfM{{\bf M}}\nc\cM{{\mathcal M}}
\nc\bfn{{\boldsymbol n}}\nc\bfN{{\bf N}}\nc\cN{{\mathcal N}}
\nc\bfo{{\boldsymbol o}}\nc\bfO{{\bf O}}\nc\cO{{\mathcal O}}
\nc\bfp{{\boldsymbol p}}\nc\bfP{{\bf P}}\nc\cP{{\mathcal P}}
\nc\bfq{{\boldsymbol q}}\nc\bfQ{{\bf Q}}\nc\cQ{{\mathcal Q}}
\nc\bfr{{\boldsymbol r}}\nc\bfR{{\bf R}}\nc\cR{{\mathcal R}}
\nc\bfs{{\boldsymbol s}}\nc\bfS{{\bf S}}\nc\cS{{\mathcal S}}
\nc\bft{{\boldsymbol t}}\nc\bfT{{\bf T}}\nc\cT{{\mathcal T}}
\nc\bfu{{\boldsymbol u}}\nc\bfU{{\bf U}}\nc\cU{{\mathcal U}}
\nc\bfv{{\boldsymbol v}}\nc\bfV{{\bf V}}\nc\cV{{\mathcal V}}
\nc\bfw{{\boldsymbol w}}\nc\bfW{{\bf W}}\nc\cW{{\mathcal W}}\nc\sW{{\mathscr W}}
\nc\bfx{{\boldsymbol x}}\nc\bfX{{\boldsymbol X}}\nc\cX{{\mathcal X}}\nc\sX{{\mathscr X}}
\nc\bfy{{\boldsymbol y}}\nc\bfY{{\boldsymbol Y}}\nc\cY{{\mathcal Y}}\nc\sY{{\mathscr Y}}
\nc\bfz{{\boldsymbol z}}\nc\bfZ{{\boldsymbol Z}}\nc\cZ{{\mathcal Z}}\nc\sZ{{\mathscr Z}}
\newcommand{\RR}{\mathbb{R}}
\newcommand{\ra}{\rightarrow}
\newcommand{\rag}{\rangle}
\newcommand{\lag}{\langle}

\newcommand{\supp}{\textup{supp}}
\newcommand{\dist}{\textup{dist}}

\global\long\def\dfn{:=}
\newcommand{\blue}{\color{black}}
\newcommand{\red}{\color{black}}

\begin{document}
\title{On the VC-Dimension of Binary Codes\thanks{This paper was presented in part at 2017 IEEE International Symposium on Information Theory.} }
\author{
Sihuang Hu\footremember{alley}{Lehrstuhl D f\"ur Mathematik, RWTH Aachen, Germany (husihuang@gmail.com). 
This work was done while S. Hu was with Department of Electrical Engineering - Systems, Tel Aviv University, Israel. 
Research supported by ERC grant no.~639573 and the Alexander von Humboldt Foundation.}
\and Nir Weinberger \footremember{trailer}{Department of Electrical Engineering--Systems, 
  Tel Aviv University, Tel Aviv, Israel (nir.wein@gmail.com, ofersha@eng.tau.ac.il).
  The work of N. Weinberger was supported by ERC grant no.~639573.
  The work of O. Shayevitz was supported by ERC grant no.~639573 and ISF grant no.~1367/14.
  }
\and Ofer Shayevitz\footrecall{trailer} 
}

\date{}
\maketitle

\begin{abstract}
  We investigate the {\red maximal} asymptotic rates of length-$n$ binary codes with VC-dimension at most $dn$ and minimum distance at least $\delta n$. 
  Two upper bounds are obtained, one as a simple  corollary of a result by Haussler and the other via a shortening approach combining {\red the} Sauer--Shelah lemma and the linear programming bound. 
  Two lower bounds are given using Gilbert--Varshamov type arguments over constant-weight and Markov-type sets. 
\end{abstract}

\section{Introduction}\label{sec:introduction}
Let $\cC\subseteq\{0,1\}^n$ be a binary code of length $n$ and rate $R=\tfrac{1}{n}\log_{2}|\cC|$. In this paper, we study the relation between the rate of the code and two fundamental properties: its minimum (Hamming) distance and its Vapnik--Chervonenkis (VC) dimension~\cite{VC1971}.  Recall that the Hamming distance between two codewords is the number of positions in which they differ; the \emph{minimum distance} of $\cC$, which is the smallest Hamming distance between any pair of codewords, plays an important role in coding theory. Recall that the projection of $\cC$ onto a coordinate set $I\subseteq [n]:=\{1,2,\ldots,n\}$, denoted $\cC|_{I}$, is the set of all possible values assigned to these coordinates by the codewords in $\cC$. The code $\cC$ is said to \emph{shatter} $I$ if $\cC|_{I}=\{0,1\}^{|I|}$. The \emph{VC-dimension} of $\cC$, which is the maximum size of a coordinate set that is \emph{shattered} by $\cC$, plays an important role in statistical learning theory and computational geometry~\cite{BEHW1989,Du1978,HaWl1987}. 

Our goal in this paper is to analyze codes of simultaneously large minimum distance and small VC-dimension. Loosely speaking, we note that fixing a rate and striving to optimize one of these properties is expected to essentially be the worst possible for the other property. Indeed, on the one hand, it is well known that random linear codes achieve the Gilbert-Varshamov bound~\cite{Gilbert1952,Varshamov1957}, which is the best known lower bound on the rate of binary codes under a minimum distance constraint, yet clearly their VC-dimension is the largest possible (attained by any information set). On the other hand, by the Sauer--Shelah lemma~\cite{Sauer1972,Shelah1972}, the VC-dimension at any given rate is essentially minimized by any Hamming ball of a suitable radius, yet clearly the minimum distance of a Hamming ball is equal to $1$, the smallest possible. These extremal observations demonstrate the tension between increasing the minimum distance and decreasing the VC-dimension. 

Besides being an interesting combinatorial problem, finding codes that have a large minimum distance as well as a small VC-dimension also admits the following coding-theoretic motivation. Suppose that a binary code $\cC$ with minimum distance $\Delta$ and VC-dimension $D$ is used over an \textit{errors and erasures} channel. Suppose there were $e$ erasures, and we are now interested in detecting whether any errors have fallen in the remaining $n-e$ coordinates. Let {\blue $t_e\in\{0,1,\ldots,n-e\}$ be the maximal number of errors that the code can guarantee to detect, and let $\pi_e\in\{0,1,\ldots,2^{n-e}\}$} the maximal number of distinct {\em error sequences} (of length $n-e$) that the code can guarantee to detect. The error detection threshold pertaining to each of these quantities is the maximal number of erasures $e$ such that the respective quantity is nonzero. {\blue If $e<\Delta-1$, then the minimum distance of the projection of $\cC$ onto the remaining $n-e$ coordinates is at least $\Delta-e>1$. Hence, the code can correct at least $\lfloor(\Delta-e)/2\rfloor$ errors and thus in this case $t_e>0$. Similarly, if $e<n-D$ then $\cC$ cannot shatter the remaining $n-e\, (>D)$ coordinates. Thus, there must be error sequences that result in vectors that are not contained in the projection of $\cC$ onto the remaining $n-e$ coordinates; such error sequences can clearly be detected, hence $\pi_e>0$.} Adopting this viewpoint, it is interesting to seek codes for which both error detection thresholds are high, namely codes with a large minimum distance and a small VC-dimension. We are interested in the maximum size of such codes.

In what follows, we consider the asymptotic formulation of the problem. 
For any\footnote{{\blue For $d\ge 1/2$ it is easy to see that the rate $R$ is always equal to $1$, which is not interesting. Therefore we limit $d$ in the interval $[0,1/2]$.}} $d,\delta\in[0,\frac{1}{2}]$, we say that a rate $R$ is $(d,\delta)$-{\it achievable} 
if for any $N$ there exists a binary code $\cC$ of length $n\geq N$, rate at least $R$, VC-dimension at most $\lfloor dn \rfloor $, and minimum distance at least $\lceil \delta n \rceil $. We are interested in characterizing $C(d,\delta)$, which we define to be the supremum of all $(d,\delta)$-achievable rates. For brevity, we assume throughout that $dn$ and $\delta n$ are integers, as this does not affect the asymptotic behavior. 

In Section~\ref{sec: upper bounds} we derive two upper bounds for $C(d,\delta)$. The first is obtained as a simple asymptotic corollary of a result by Haussler~\cite{Haussler1995}, and the second is derived via a shortening approach that combines the Sauer--Shelah lemma~\cite{Sauer1972,Shelah1972} (controlling the VC-dimension) and the linear programming bound~\cite{MRRW1977} (controlling the minimum distance). In Section~\ref{sec:LowerBounds} we present two lower bounds for $C(d,\delta)$. Both these bounds are obtained via GV-type arguments (controlling the minimum  distance) applied to constant-weight and Markov-type sets respectively (whose structure controls the VC-dimension).  

\section{Upper Bounds}\label{sec: upper bounds}
We first briefly review upper bounds on $C(d,\delta)$ that can be easily deduced from known results. To begin, one
can clearly ignore either the minimal distance constraint or the VC-dimension
constraint. 

When accounting only for the minimal distance constraint,
the best known upper bound is the \emph{second MRRW bound}
given by McEliece, Rodemich, Rumsey, and Welch~\cite{MRRW1977} as follows:
\[
  R_{LP}(\delta)\dfn\min_{0\le u\le 1-2\delta}\{1+g(u^2)-g(u^2+2\delta u+2\delta)\}
\]
with $g(x):=h( (1-\sqrt{1-x})/2)$.
{\blue Here and throughout this paper we define $h(x)=-x\log_2(x)-(1-x)\log_2(1-x)$ to be the binary entropy function.}
The following is direct.
\begin{lemma}\label{lem: MRRW}
$C(d,\delta)\leq R_{LP}(\delta).$
\end{lemma}

When accounting only for the VC-dimension constraint, the size of
a code ${\cal C}$ with VC-dimension $dn$ can be upper bounded by the Sauer\textendash Shelah
lemma~\cite{Sauer1972,Shelah1972}
\begin{align} \label{eq: PSS}
|{\cal C}|\leq\sum_{i=0}^{dn}{n \choose i}
\end{align}
and so the following is evident.
\begin{lemma}\label{lem: PSS bound}
$C(d,\delta)\leq h(d).$
\end{lemma}
In~\cite{Haussler1995} Haussler directly addressed the problem of bounding the size of codes with restricted minimal distance and VC-dimension. In his setting, the VC-dimension is a bounded constant.
However, from the results there the following bound on $C(d,\delta)$ can still be deduced. For a number $a\ge0$ we define $\left\langle a\right\rangle \dfn\min(a,\frac{1}{2})$. 
  For a code $\cC$ we define the \emph{unit distance graph } UD($\cC$) whose vertex set is all codewords in $\cC$ and two codewords $\bfx,\bfy$ are adjacent if their Hamming distance $\dist(\bfx,\bfy)=1$.

\begin{lemma}[{Corollary to~\cite[Theorem 1]{Haussler1995}}]
\label{lem: Haussler's bound}
\[
C(d,\delta)\leq 
\frac{2d}{\delta+2d}\cdot h\left(\left\langle \frac{\delta+2d}{2}\right\rangle \right).
\]
\end{lemma}
\begin{proof}
  Let $\cC$ be a length-$n$ binary code with VC-dimension at most $dn$ and minimum distance $\delta n$. 
  Suppose $0\le s\le 1$. {\blue We choose a random subset $I\subseteq[n]:=\{1,2,\dots,n\}$ of size $sn$ uniformly.}
  For each codeword $\bfu\in\cC|_I$, we define its weight $w(\bfu)$ as the number of codewords in $\cC$ such that its projection on $I$ is equal to $\bfu$. Let $E$ be the edge set of the unit distance graph UD($\cC|_I$), and define the weight of an edge $e=\{\bfu,\bfv\}$ as $w(e)=\min\{w(\bfu),w(\bfv)\}$. Put $W=\sum_{e\in
  E}w(e)$, and note that $W$ is a random variable depending on the random choice of $I$. The bound follows by estimating $\bfE[W]$, the expectation of $W$, in two ways.
  First, we claim that for any $I\subset[n]$, 
  \begin{align}\label{eq:upper}
    W\le 2dn|\cC|.
  \end{align}
  On the other hand, we can bound $\bfE[W]$ from below:
  \begin{align}\label{eq:lower}
    \bfE[W]\ge\frac{sn\cdot\delta n}{n-sn+1}\left(|\cC|-\sum_{i=0}^{dn}{sn\choose i}\right).
  \end{align}
  (Please refer to~\cite[Lemma 5.14]{Ma2010} for the proof of~\eqref{eq:upper} and~\eqref{eq:lower}.) Thus we have
  \begin{align*}
    \left(((\delta+2d)s-2d)-\frac{2d}{n}\right)|\cC|\le s\delta\sum_{i=0}^{dn}{sn\choose i}.
  \end{align*}
For any $s>\frac{2d}{\delta+2d}$ and sufficient large $n$, we can get $|\cC|=O(\sum_{i=0}^{dn}{sn\choose i})$, and hence $C(d,\delta)\le s\cdot h(\langle{d}/{s}\rangle).$ The result follows directly.
\end{proof}

We shall next combine Lemma~\ref{lem: MRRW} and Lemma~\ref{lem: PSS bound} to obtain an improved upper bound.
{\blue Throughout this paper, we define $0/0=0$.}
\begin{theorem}\label{thm: Upper bound}
\begin{align*}
C(d,\delta)\leq\min_{0\leq s\leq1-2\delta}\left\{ s\cdot h\left(\left\langle \frac{d}{s}\right\rangle \right)+(1-s)R_{LP}\left(\frac{\delta}{1-s}\right)\right\}.
\end{align*}
\end{theorem}
\begin{proof}
Let $\cC$ be a length-$n$ binary code with VC-dimension at most $dn$ and minimum distance $\delta n$.
Choose $s\in[0,1-2\delta]$, and consider the projection of ${\cal C}$ on
$[sn]=\{1,2,\dots,sn\}$. 
Of course the VC-dimension of $\cC|_{[sn]}$ is also at most $dn$, and so its rate can be bounded by Lemma~\ref{lem: PSS bound}.
For any given prefix $\bfu\in\cC|_{[sn]}$,
we denote the set of its possible suffixes by $\mathcal{Z}(\bfu)\subset\{0,1\}^{(1-s)n}$, i.e., for any $\bfv\in\cZ(\bfu)$ there exists a codeword $\bfx\in\cC$ such that $\bfx$ is the concatenation of $\bfu$ and $\bfv$.
Clearly, $\mathcal{Z}(\bfu)$ is a code of length $(1-s)n$
and minimal distance $\delta n$, and so its rate can be bounded by 
the second MRRW bound.
Then our result follows from
\[
  |{\cal C}|
  = \sum_{\bfu\in\cC|_{[sn]}}\left|\mathcal{Z}(\bfu)\right|
  \leq \Big|{\cal C}|_{[sn]}\Big|\cdot\max_{\bfu\in\cC|_{[sn]}}\left|\mathcal{Z}(\bfu)\right|.
\]
\end{proof}

\section{Lower Bounds}\label{sec:LowerBounds}
A general procedure to obtain lower bounds on $C(d,\delta)$ is the following.
\begin{enumerate}[(i)]
  \item 
    Pick some subset $S$ of the Hamming cube $\{0,1\}^n$ that has some ``nice'' structure.
  \item
  Compute a generalized GV bound for subset $S$, namely a lower bound on the size of the largest code of minimum distance at least $\delta n$ where all codewords belong to $S$.
\item
Find an upper bound for the VC-dimension $dn$ of any subset of $S$ that has minimum distance at least $\delta n$.
\item
Combine the bounds (ii)-(iii).
\end{enumerate}

In the following two subsections, we will show two ways to choose ``nice'' subsets of the Hamming cube and calculate the corresponding bounds.

\subsection{Constant Weight Codes}\label{subsec:CWC}
Here we choose subset $S$ to be the collection of all codewords with some constant weight.

\begin{lemma}\label{VCDofCWC}
Suppose $\delta\in[0,\frac{1}{2}]$ and $w\in[0,1]$. Let $\cC$ be a binary code of length $n$, constant weight $wn$, and minimum distance $\delta n$. Then the VC-dimension of $\cC$ is at most $(w-\delta/{2})n+1$.
\end{lemma}
\begin{proof}
Suppose the VC-dimension of $\cC$ is $dn$. Without loss of generality, we assume that the first $dn$ coordinates are shattered. Then there exist two codewords $\bfx=x_1x_2\cdots x_n$ and $\bfy=y_1y_2\cdots y_n$ such that $x_i=1$ for $1\le i\le dn$ and $y_i=1$ for $1\le i\le dn-1$ and $y_{dn}=0$. Hence $|\supp(\bfx)\cap \supp(\bfy)|\ge dn-1$. On the other hand, $\dist(\bfx,\bfy)=2wn-2|\supp(\bfx)\cap \supp(\bfy)|$, which is at least $\delta n$. Therefore
$\delta n\le 2wn-2|\supp(\bfx)\cap \supp(\bfy)|\le 2wn-2(dn-1)$. This proves the result.
\end{proof}

Let $A(n,\delta n,wn)$ denote the maximum size of length-$n$ binary code with constant weight $wn$ and minimum distance $\delta n$. The following GV-type bound is well-known.
\begin{lemma}
\begin{align}\label{CWCGV}
  A(n, \delta n, wn) \ge \frac{{n \choose wn}}{\sum_{i=0}^{\delta n/2-1}{wn\choose i}{n-wn\choose i}}.
\end{align}
\end{lemma}

Now we are ready to state our first lower bound for $C(d,\delta)$.

\begin{theorem} \label{CWCBound}
  Let $d,\delta\in[0,\frac{1}{2}]$, and let $w=d+\frac{\delta}{2}$. Then
  \begin{align*}
    C(d,\delta)\ge
    \begin{cases}
      h(w)-{\displaystyle\max_{0\le x\le\delta/2}}\Big[w\,h\Big(\frac{x}{w}\Big) +(1-w)h\Big(\frac{x}{1-w}\Big)\Big]& \textup{if } w<\frac{1}{2}\\
      1-h(\delta) &\textup{otherwise}.
    \end{cases}
\end{align*}
\end{theorem}
\begin{proof}
  If $w<\frac{1}{2}$, plug it into~\eqref{CWCGV} and take the asymptotic form, then the result follows directly from Lemma~\ref{VCDofCWC}. If $w\ge\frac{1}{2}$ then set $w=\frac{1}{2}$ in~\eqref{CWCGV} {\blue which maximizes the lower bound}.
\end{proof}

\subsection{Markov Type}\label{subsec:Markov}
For a binary codeword $\bfx=x_1x_2\cdots x_n\in\{0,1\}^n$, 
the number of \emph{switches} of $\bfx$ is equal to $|\{i: 1\le i\le n-1, x_i\oplus x_{i+1}=1\}|$, 
  {\blue that is the number of length-$2$ consecutive subsequence $01$ or $10$.}
  (Here $\oplus$ is the XOR operation.)
Now we present another lower bound for $C(d,\delta)$ based on the following observation. 

\begin{fact}
Let $S$ be the collection of all codewords in the Hamming cube $\{0,1\}^n$ that has at most $dn$ switches. 
Then the VC-dimension of $S$ or any subset of $S$ is at most $dn+1$.  
\end{fact}
{\blue
\begin{proof}
  Let $I$ be any $dn+2$ coordinates. 
  Let $\bfc$ be a length-($dn+2$) vector such that $c_i=0$ for odd $i\in\{1,2,\dots,dn+2\}$ and  
  $c_i=1$ for even $i\in\{1,2,\dots,dn+2\}$. Then the number of switches of $\bfc$ is $dn+1$.
  Hence the projection of $S$ onto these coordinates $S|_{I}$ does not contain $\bfc$, therefore
  $S$ does not shatter $I$. This concludes our proof.
\end{proof}
}

We refer to an $(S,M,\delta n)$-code as a subset of $S$ with size $M$ and minimum distance at least $\delta n$. We will prove a GV-type bound for such $(S,M,\delta n)$-codes, and thus get a lower bound for $C(d,\delta)$. 
Our proof relies on a generalized GV bound provided by Kolesnik and Krachkovsky~\cite{KolesnikKrachkovsky1991}, and follows the same line of reasoning as in Sections III-V of~\cite{MarcusRoth1992}, where Marcus and Roth developed an improved GV bound for constrained systems based on stationary Markov chains. 

\begin{lemma}\cite[Lemma 1]{KolesnikKrachkovsky1991}
  \label{GeneralizedGV}
  Let $S$ be a subset of $\{0,1\}^n$. Then there exists an $(S,M,\delta n)$-code such that
  $$M \ge \frac{|S|^2}{4|\cB_S(\delta n-1)|}$$
where
$$\cB_S(\delta n-1):=\{(\bfw,\bfw')\in S\times S: \dist(\bfw,\bfw')\le \delta n-1\}.$$
\end{lemma}

In order to compute our lower bound, we shall consider stationary Markov chains on graphs. A \emph{labeled graph} $G=(V_G,E_G,L_G)$ is a finite \emph{directed} graph with vertices $V_G$, edges $E_G$, and a labeling $L_G:E_G\ra \Sigma$ for some finite alphabet $\Sigma$. For any vertex $u$, the set of outgoing edges from $u$ is denoted by $E^+_G(u)$, and the set of incoming edges to $u$ is $E^-_G(u)$. 
A graph $G$ is called \emph{irreducible} if there is a path in each direction between each pair of vertices of the graph. The greatest common divisor of the lengths of cycles of a graph $G$ is called the \emph{period} of G. An irreducible graph $G$ with period $1$ is called \emph{primitive}.
A \emph{stationary Markov chain} on a finite directed graph $G$ is a function $P:E_G\ra [0,1]$ such that
\begin{enumerate}[(i)]
  \item $\sum_{e\in E_G} P(e)=1$;
  \item $\sum_{e\in E_G^+(u)}P(e)=\sum_{e\in E_G^-(u)}P(e)$ for every $u\in V_G$. 
\end{enumerate}

{\blue Evidently, $P(e)$ represents the probability that the chain will make a transition along the edge $e$.} We denote by $\cM(G)$ the set of all stationary Markov chains on $G$.
For a stationary Markov chain $P\in\cM(G)$, we introduce two dummy random variables $X,Y$ such that their joint distribution is defined by
\begin{align*}
  \Pr\{X=u, Y=v\}=
    \begin{cases}
      P((u,v)) &\text{if } (u,v)\in E_G\\
      0        &\text{otherwise.}
    \end{cases}
\end{align*}
Then the condition (ii) amounts to saying that the marginal distributions of $X$ and $Y$ are equal.


For a stationary Markov chain $P\in\cM(G)$ and a function $f:E_G\ra \RR^k$, we denote by $\bfE_P(f)$ the expected value of $f$ with respect to $P$, that is,
$$ \bfE_P(f):=\sum_{e\in E_G}P(e)f(e). $$
Fix a vertex $u$, and let $\Gamma_n(G)$ denote the set of all cycles in $G$ of length $n$ that start and end at $u$. For a cycle $\gamma=e_1e_2\dots e_n\in\Gamma_n(G)$, let $P_{\gamma}$ denote the stationary Markov chain defined by
$$ P_{\gamma}(e):=\frac{1}{n}|\{i\in \{1,2,\dots,n\}: e_i=e\}|. $$
We refer to $P_{\gamma}$ as the \emph{empirical distribution} of the cycle $\gamma$, and to
$$ \bfE_{P_{\gamma}}(f) = \sum_{e\in E_G}P_{\gamma}(e)f(e) $$
as the \emph{empirical average} of $f$ on the cycle $\gamma$.
(Note that the empirical distribution $P_{\gamma}$ is closely related to the so-called ``second-order type'' of sequence $L_G(e_1)L_G(e_2)\cdots L_G(e_n)$.)
For a subset $U\subset \RR^k$, let $\cM(G;f,U)$ denote the set of all stationary Markov chains $P$ on $G$ such that $\bfE_{P}(f)\in U$, and let 
$$ \Gamma_n(G;f,U):=\{\gamma\in\Gamma_n(G): \bfE_{P_{\gamma}}(f)\in U\}.$$ 

The following lemma is a consequence of well-known results on second-order types of Markov chains, cf. Boza~\cite{Boza1971}, Davisson, Longo, Sgarro~\cite{DaLoSg1981}, Natarajan~\cite{Na1985}, Csisz\'ar, Cover, Choi~\cite{CsCoCh1987}, and Csisz\'ar~\cite{Cs1998}. (Throughout this paper, the base of the logarithm is $|\Sigma|$.)

\begin{lemma}\cite[Lemma 2]{MarcusRoth1992}
  \label{MarkovType}
  Let $G$ be a primitive graph and $f:E_G\ra \RR^k$ be a function on the edges of $G$. Let $U$ be an open and nonempty subset of $\RR^k$. Then
  \begin{align*}
    \lim_{n\ra\infty}\frac{1}{n}\log|\Gamma_n(G;f,U)|=\sup_{P\in\cM(G;f,U)}H_P(Y|X).
  \end{align*}
\end{lemma}

\begin{figure}[!t]
  \centering
  \includegraphics[width=3 in]{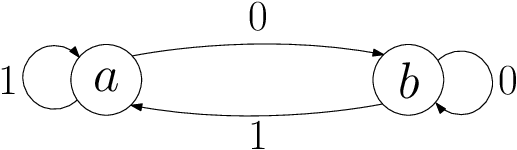}
  \caption{labeled graph $G$ over $\Sigma=\{0,1\}$}
  \label{fig:LabeledGraph}
\end{figure}

Hereafter we will consider the labeled graph $G$ over alphabet $\Sigma=\{0,1\}$ depicted in Figure~\ref{fig:LabeledGraph}. The labeling $L_G$ is defined by $L_G((a,b))=L_G((b,b))=0$ and $L_G((b,a))=L_G((a,a))=1$. On the other hand, the function $f:E_G\ra\RR$ is defined by $f((a,a))=f((b,b))=0$ and $f((a,b))=f((b,a))=1$. Then we can verify the following. 

\begin{fact}
For a cycle $\gamma=e_1e_2\cdots e_n\in\Gamma_n(G)$, the value $nE_{P_{\gamma}}(f)-f(e_1)$ is equal to the number of switches of the corresponding binary sequence $L_G(e_1)L_G(e_2)\cdots L_G(e_n)$.
\end{fact}

Now we come to our second lower bound for $C(d,\delta)$. We will consider the subset
\begin{align*}
S_n(d)=S_n([0,d]):= \{L_G(e_1)L_G(e_2)\cdots L_G(e_n): e_1e_2\cdots e_n\in\Gamma_n(G;f,[0,d])\}.
\end{align*}
By definition, for any $\bfx\in S_n(d)$ its number of switches is at most $dn$. 

In order to use Lemma~\ref{GeneralizedGV}, we introduce the graph $G\times G$ whose vertex set is $V_{G\times G}=V_G\times V_G=\{\langle u,u'\rangle: u,u'\in V_G\}$ and edge set is $E_{G\times G}=E_G\times E_G=\{\langle e,e'\rangle: e,e'\in E_G\}$. Given the function $f$ defined on the edges of $G$, we define two functions $f^{(1)}$ and $f^{(2)}$ on $E_{G\times G}$ by 
$$f^{(1)}(\lag e,e'\rag)=f(e),\quad f^{(2)}(\lag e,e'\rag)=f(e')$$
and a function $\Delta:E_{G\times G}\ra\RR$ by
$$\Delta(\lag e,e'\rag)=
        \begin{cases}
          1 & \text{if } L_G(e)\neq L_G(e')\\
          0 & \text{otherwise.}
        \end{cases}
        $$
{\blue Note that the function $\Delta$ is used to count the Hamming distance between two binary sequences.}
We collect $f^{(1)},f^{(2)}$ and $\Delta$ to define a function $\varphi: E_{G\times G}\ra \RR^3$ by $\varphi=[f^{(1)},f^{(2)},\Delta]$.
For a subset $U\subset [0,1]$ we set
\begin{align*}
\cF(U) &:=\sup_{P\in\cM(G; f, U)} H_P(Y|X),\\
\cG(U,\delta) &:=\sup_{Q\in\cM(G\times G;\varphi,U\times U\times [0,\delta))}H_Q(Y|X).
\end{align*}
In particular, we use $\cF(p)$ and $\cG(p,\delta)$ as short-hand notations for $\cF(\{p\})$ and $\cG(\{p\},\delta)$ respectively, where $0\le p\le1$.
Set
\begin{align*}
  \begin{split}
&R_{MA}(d,\delta)
:=\sup_{p\in[0,d]}\{2\cF(p)-\cG(p,\delta)\}\\
=&\sup_{p\in[0,d]}\Big\{2\sup_{\substack{P\in\cM(G):\\\bfE_P(f)=p}}H_P(Y|X)-\sup_{\substack{Q\in\cM(G\times G):\\ \bfE_Q(f^{(i)})=p,\,i=1,2\\ \bfE_Q(\Delta)\in[0,\delta)}}H_Q(Y|X)\Big\}.
\end{split}
\end{align*}

\begin{lemma}\label{MarkovLB}
  There exist $(S_n(d),M,\delta n)$-codes satisfying
  $$\frac{\log M}{n}\ge R_{MA}(d,\delta)-o(1).$$
\end{lemma}
\begin{proof}
  For $p\in[0,d]$ and $\varepsilon>0$, let $U_{p,\varepsilon}=(p-\varepsilon,p+\varepsilon)$, 
  \begin{align*}
     S_n(U_{p,\varepsilon}):=\{L_G(e_1)L_G(e_2)\cdots L_G(e_n): e_1e_2\cdots e_n\in\Gamma_n(G;f,U_{p,\varepsilon})\},
  \end{align*}
 and
  \begin{align*}
    \cB_{S_n(U_{p,\varepsilon})}(\delta n-1):=\{(\bfw,\bfw')\in S_n(U_{p,\varepsilon})\times S_n(U_{p,\varepsilon}): \dist(\bfw,\bfw')\le \delta n-1\}.
  \end{align*}
  By Lemma~\ref{MarkovType},
  \begin{align*}
    \lim_{n\rightarrow\infty}\frac{1}{n}\log{|S_n(U_{p,\varepsilon})|} =\lim_{n\rightarrow\infty}\frac{1}{n}\log{|\Gamma_n(G;f,U_{p,\varepsilon})|}=\cF(U_{p,\varepsilon}),
  \end{align*}
  and
  \begin{align*}
      \lim_{n\rightarrow\infty}\frac{1}{n}\log{|\cB_{S_n(U_{p,\varepsilon})}(\delta n-1)|}
    =\lim_{n\rightarrow\infty}\frac{1}{n}\log{|\Gamma_n(G\times G;\varphi;U_{p,\varepsilon}\times U_{p,\varepsilon}\times[0,\delta))|}
   =\cG(U_{p,\varepsilon},\delta).
  \end{align*}
  Note that both $H_P(Y|X)$ and $\bfE_{P}(f)$ are continuous in $P$.
  {\blue So if we let $\varepsilon\rightarrow0$, then by Lemma~\ref{GeneralizedGV} there exist $(S_n(d),M,\delta n)$-codes satisfying
  \begin{align*}
  \frac{\log M}{n}\ge 2\cF(p)-\cG(p,\delta)-o(1).
  \end{align*}
  }
Then our result follows.
\end{proof}


\begin{theorem}\label{thm:MK}
  $ C(d,\delta)\ge R_{MA}(d,\delta).$
\end{theorem}
\begin{proof}
  This follows from Lemma~\ref{MarkovLB} and the fact that any $(S_n(d),M,\delta n)$ code has VC-dimension at most $dn+1$.
\end{proof}

Using convex duality we can compute $R_{MA}(d,\delta)$ through an unconstrained optimization problem with convex objective function as follows.
For a function $f:E_{G}\ra \RR^k$, let $A_{G;f}(\bfx),\bfx\in\RR^k$, be the matrix function indexed by the states of $G$ with entries
$$[A_{G;f}(\bfx)]_{u,v}=
\begin{cases}
  2^{-\bfx\cdot f((u,v))} &\text{if } (u,v)\in E_{G}\\
  0 &\text{otherwise},
\end{cases}
$$
and let $\lambda_{G;f}(\bfx)$ denote the spectral radius of $A_{G;f}(\bfx)$.
{\blue (Here the $\cdot$ operator in the exponent is the inner product of two vectors.)}
%
Recall the definitions of $f,f^{(1)},f^{(2)}, \Delta, \varphi$, and define $\varphi'=[f^{(1)}+f^{(2)},\Delta]: E_{G\times G}\ra \RR^2.$
%
Let $G$ be the graph of Figure 1. Then
\begin{align*}
  A_{G;f}(x) = \kbordermatrix{
    & a & b \\
  a & 1 & 2^{-x}\\
  b &2^{-x} & 1
  }
\end{align*}
and
\begin{align*}
  A_{G\times G;\varphi'}(x,z)=\kbordermatrix{
    & \lag a,a\rag  & \lag a,b\rag  & \lag b,a\rag  & \lag b,b\rag  \\
    \lag a,a\rag  &  1 & 2^{-x-z} & 2^{-x-z} & 2^{-2x} \\
    \lag a,b\rag  &  2^{-x} & 2^{-z} & 2^{-2x-z} & 2^{-x} \\
    \lag b,a\rag  &  2^{-x} & 2^{-2x-z} & 2^{-z} & 2^{-x}\\
    \lag b,b\rag  &  2^{-2x} & 2^{-x-z} & 2^{-x-z} & 1
  }.
\end{align*}
Through direct computations, we have
  $\lambda_{G;f}(x)=2^{-x}+1,$
and
\begin{align*}
  \begin{split}
   \lambda_{G\times G;\varphi'}(x,z)&= \frac{1}{2} \Big((4^{-x}+1)(2^{-z}+1)+\\
   &\hspace{-2cm}\sqrt{(4^{-x}+1)^2 4^{-z}-2(16^{-x}-6\cdot4^{-x}+1)2^{-z}+(4^{-x}+1)^2}\Big).
 \end{split}
\end{align*}

From the well-known results in convex duality principle, we can obtain the following.
Similar results are also obtained in~\cite{JuHo1984,MaTu1990}.

\begin{lemma}\cite[Lemma 5]{MarcusRoth1992}
  \label{lem: optidual}
  Let $G$ be a graph and let $f:E_{G}\to\RR^k, g:E_G\to\RR^l$ be functions on the edges of $G$. Set $\phi=[f,g]:E_{G}\to\RR^{k+l}$. Then for any $\bfr\in\RR^k$ and $\bfs\in\RR^l$,
  \begin{align*}
    \sup_{\substack{P\in\cM(G):\\\bfE_P(f)=\bfr\\\bfE_P(g)\le\bfs}}H_P(Y|X)=\inf_{\substack{\bfx\in\RR^k\\\bfz\in\RR_{\ge0}^{l}}}\{\bfx\cdot\bfr+\bfz\cdot\bfs+\log{\lambda_{G;\phi}(\bfx,\bfz)}\}.
  \end{align*}
\end{lemma}

\begin{theorem}
  \begin{align*}
     R_{MA}(d,\delta)
    = \sup_{p\in [0,d]}\Big\{2\,h(p)-\inf_{\substack{x\in\RR\\z\in\RR_{\ge0}}}\{2px+\delta z+\log{\lambda_{G\times G;\varphi'}(x,z)}\}\Big\}.
  \end{align*}
\end{theorem}
\begin{proof}
  Applying Lemma~\ref{lem: optidual} to compute $\cF(p)$, we have
  \begin{align*}
    \cF(p)
    &=\sup_{\substack{P\in\cM(G):\\\bfE_P(f)=p}}H_P(Y|X)\\
    &=\inf_{x\in\RR}\{px+\log{\lambda_{G;f}(x)}\}\\
    &=\inf_{x\in\RR}\{px+\log{(2^{-x}+1)}\}\\
    &=h(p).
  \end{align*}
  Similarly, we have
  \begin{align*}
    \cG(p,\delta)
    &=\sup_{\substack{Q\in\cM(G\times G):\\ \bfE_Q(f^{(i)})=p,\,i=1,2\\ \bfE_Q(\Delta)\in[0,\delta)}}H_Q(Y|X)\\
    &=\inf_{\substack{x,y\in\RR\\z\in\RR_{\ge0}}}\{px+py+\delta z+\log{\lambda_{G\times G;\varphi}(x,y,z)}\}\\
    &\le\inf_{\substack{x\in\RR\\z\in\RR_{\ge0}}}\{2px+\delta z+\log{\lambda_{G\times G;\varphi}(x,x,z)}\}\\
    &=\inf_{\substack{x\in\RR\\z\in\RR_{\ge0}}}\{2px+\delta z+\log{\lambda_{G\times G;\varphi'}(x,z)}\}.
  \end{align*}
  On the other hand, for $\varepsilon>0$, choose some point $(x',y',z')$ such that 
  $$px'+py'+\delta z'+\log{\lambda_{G\times G;\varphi}(x',y',z')} \le 
   \inf_{\substack{x,y\in\RR\\z\in\RR_{\ge0}}}\{px+py+\delta z+\log{\lambda_{G\times G;\varphi}(x,y,z)}\}+\varepsilon,$$
  and let $\bar{x}=(x'+y')/2$. Note that $\lambda_{G\times G;\varphi}(x,y,z)=\lambda_{G\times G;\varphi}(y,x,z)$ and  
  the function $\log{\lambda_{G\times G;\varphi}(x,y,z)}$ is convex (see~\cite[Remark 2]{MarcusRoth1992}). 
  Thus
  \begin{align*}
    &\ px'+py'+\delta z'+\log{\lambda_{G\times G;\varphi}(x',y',z')}\\
    =&\ 2p\bar{x}+\delta z'+\log{\lambda_{G\times G;\varphi}(x',y',z')}\\
    \ge&\ 2p\bar{x}+\delta z'+ \log{\lambda_{G\times G;\varphi}(\bar{x},\bar{x},z')},
  \end{align*}
  and $ \cG(p,\delta)=\inf_{\substack{x\in\RR\\z\in\RR_{\ge0}}}\{2px+\delta z+\log{\lambda_{G\times G;\varphi'}(x,z)}\}$.
  This concludes our proof.
  \end{proof}

\section{Examples}
\begin{example}{\rm
We plot the bounds for $d=\frac{1}{4}$ and $\frac{1}{16}$ in Fig. 2. 
Note that all these bounds intersect at $R=h(d)$ when $\delta=0$; and our shortening upper bound (Thm.~\ref{thm: Upper bound}) is always better than the second MRRW bound (hence we do not plot it here).
As we can see, for $d=\frac{1}{4}$ our shortening upper bound (Thm.~\ref{thm: Upper bound}) is always better than Haussler's upper bound (Lem.~\ref{lem: Haussler's bound}), and the constant weight lower bound (Thm.~\ref{CWCBound}) is always better than the Markov type lower bound (Thm.~\ref{thm:MK}). For $d=\frac{1}{16}$, the performance of these bounds are quite different. 
\begin{figure}[!t]
  \centering
  \includegraphics[width=3 in]{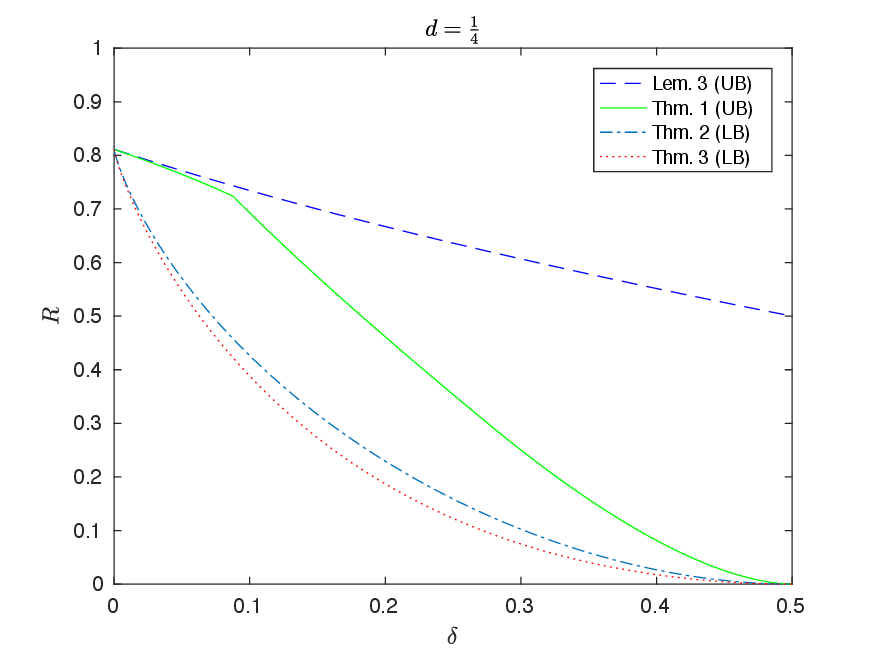}
  \includegraphics[width=3 in]{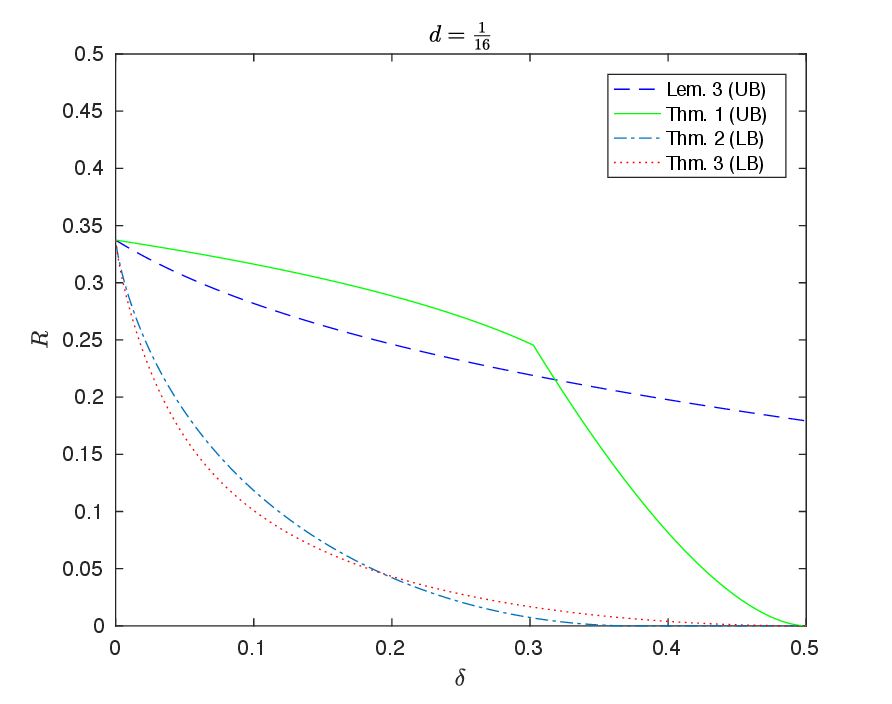}
  \caption{Bounds for $d=\frac{1}{4}$ and $d=\frac{1}{16}$}
  \label{fig:Bounds1}
\end{figure}
}
\end{example}

\begin{example}{\rm
We plot the bounds for $\delta=\frac{1}{4}$ and $\frac{1}{16}$ in Fig. 3. 
\begin{figure}[!t]
  \centering
  \includegraphics[width=3 in]{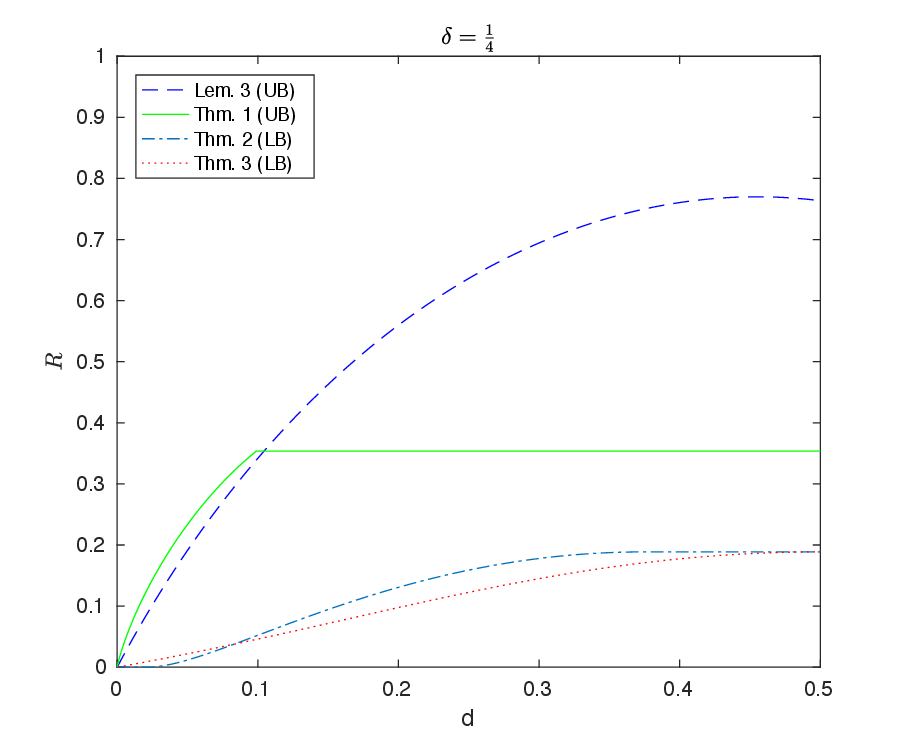}
  \includegraphics[width=3 in]{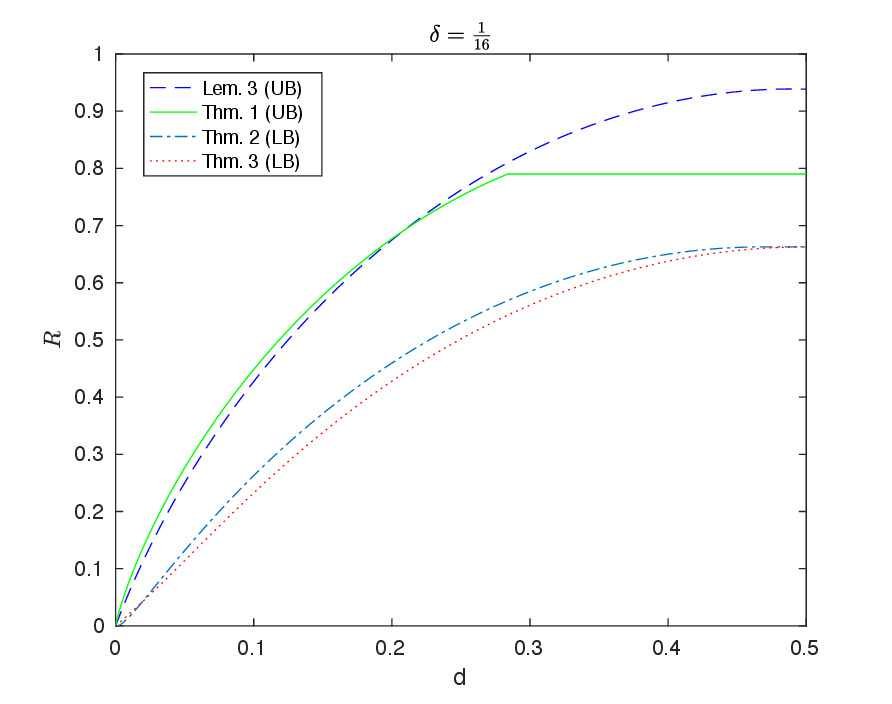}
  \caption{Bounds for $\delta=\frac{1}{4}$ and $\delta=\frac{1}{16}$}
  \label{fig:Bounds2}
\end{figure}
}
\end{example}

\begin{remark}\label{ronny}{\rm
    Similarly as in~\cite{MarcusRoth1992}, we can slightly improve the lower bounds by considering subsets $T$ of our chosen set $S$. For example, when $d=1/16$ and $\delta=0.1927$, both Theorem~\ref{CWCBound} and Theorem~\ref{thm:MK} give that $C(d,\delta)\ge0.046$. On the other hand, let $T$ be the collection of all codewords in the Hamming cube $\{0.1\}^n$ that has weight $0.5n$ and at most $1/16n$ switches, then the generalized GV bound for subset $T$ shows that $C(d,\delta)\ge0.0461$. 
}
\end{remark}

\section{Discussion}
In this paper, we have studied the maximal size of a binary code with a given minimum distance and a given VC dimension. We gave two lower bounds, based on the idea of random GV-type constructions inside structured sets (Hamming balls, Markov types) in a way that simultaneously controls the minimum distance and the VC dimension. It may be interesting to consider other structured sets in order to improve the bounds, or to come up with a different method of construction. 

Our weakest point is arguably the upper bound, which unlike the lower bounds, was derived by treating the problem of minimum distance and VC dimension separately. It stands to reason that a different argument that simultaneously controls both quantities could improve our bound. However, so far we have been unable to come up with such an argument. One reasonable line of attack could be to take the VC dimension constraint into consideration as part of an LP-type argument. However, the VC dimension constraint is global, and our attempts to embed it in the more local LP-type approach have not been fruitful. Another direction to consider is a blow-up argument: Given a code with minimum distance $\delta$, we blow-up the code to include parts of the Hamming balls of radius $\delta/2$ around each codeword. If this can be done in a controlled way such that the increase in the VC dimension can be accounted for, then the Sauer--Shelah lemma can be applied to the blown-up code. This currently appears to be difficult. 
Lastly, it would be interesting to see if a suitable shifting argument that somehow keeps the minimum distance in check can be used, to yield a bound in the spirit of the Sauer--Shelah lemma.

\section*{Acknowledgement}
We would like to thank Ronny Roth for his helpful comments on Remark~\ref{ronny}.


\bibliographystyle{siamplain}
\bibliography{VC_dimension}
\end{document}